\documentclass[epjCONF]{svjour-mod}
\pdfoutput=1
\usepackage{graphicx}
\usepackage[varg]{txfonts} % Times fonts
\usepackage[latin1]{inputenc}

\usepackage{atlasphysics}
\usepackage[hyperindex,breaklinks]{hyperref} 
\session-title{International Conference on New Frontiers in Physics 2012}
\begin{document}
\title{Highlights from SUSY searches with ATLAS}
\author{Vasiliki A.\ Mitsou\thanks{\email{vasiliki.mitsou@ific.uv.es}} \\ On behalf of the ATLAS Collaboration}
\institute{Instituto de F\'isica Corpuscular (IFIC), CSIC -- Universidad de Valencia, \\ Parc 
Cient\'ific, C/ Catedr\'atico Jos\'e Beltr\'an 2, 
E-46980 Paterna (Valencia), Spain}
\abstract{
Supersymmetry (SUSY) is one of the most relevant scenarios of new physics searched by the ATLAS experiment at the CERN Large Hadron Collider. In this write-up the principal search strategies employed by ATLAS are outlined and the most recent results for analyses targeting SUSY discovery are discussed. A wide range of signatures is covered motivated by various theoretical scenarios and topologies: strong production, third-generation fermions, long-lived particles and $R$-parity violation, among others. The results are based on up to $\sim5~\ifb$ of data recorded during 2010 -- 2011 at $\rts = 7\TeV$ centre-of-mass energy by the ATLAS experiment at the LHC.
}
\maketitle
\section{Introduction}
\label{sec:intro}
Supersymmetry (SUSY)~\cite{Nilles:1983ge} is an extension of the Standard Model (SM) which assigns to each SM field a superpartner field with a spin differing by a half unit. SUSY provides elegant solutions to several open issues in the SM, such as the hierarchy problem, the identity of dark matter, and grand unification.

SUSY searches in collider experiments typically focus on events with high transverse missing energy (\MET) which can arise from (weakly interacting) Lightest Supersymmetric Particles (LSPs), in the case of $R$-parity conserving SUSY, or from neutrinos produced in LSP decays, when $R$-parity is broken. Hence, the event selection criteria of inclusive channels are based on large \MET, no or few leptons ($e$, $\mu$), many jets and/or $b$-jets, $\tau$-leptons and photons. The exact sets of cuts (``signal regions'', SRs) are a compromise between the necessity to suppress events coming from known SM processes while maintaining sufficient number of surviving SUSY events. Typical SM backgrounds are top-quark production  ---including single-top---, $W$/$Z$ in association with jets, dibosons and QCD multi-jet events. These are estimated using semi- or fully data-driven techniques. Although the various analyses are motivated and optimised for a specific SUSY scenario, the interpretation of the results are extended to various SUSY models or topologies. 

A brief summary of recent results (as of June~2012) on searches for SUSY with and without $R$-parity conservation and for long-lived massive superpartners is presented. The reported results are based on up to 4.7~\ifb\ of data from $pp$ collisions at a center-of-mass energy of $\rts = 7\TeV$ recorded in 2010 -- 2011 by ATLAS~\cite{Aad:2008zzm} at the Large Hadron Collider (LHC)~\cite{Evans:2008zzb}.

\section{Inclusive channels}
\label{sec:RPC}

Analyses exploring $R$-parity conserving (RPC) SUSY models are currently divided into inclusive searches for: (a) squarks and gluinos, (b) third-generation fermions, and (c) electroweak production ($\tilde{\chi}^0$, $\tilde{\chi}^{\pm}$, $\tilde{\ell}$). Recent results from each category of ATLAS searches are presented in this Section. It is stressed that, although these searches are designed to look for RPC SUSY, interpretation in terms of $R$-parity violating models is also possible (cf.\ Sec.~\ref{sec:RPV}).

\subsection{Squarks and gluinos}
\label{sec:strong}

Strong SUSY production is searched in events with large jet multiplicities and large missing transverse momentum, with and without leptons. Various channels fall into this class of searches; here two cases are highlighted: the 0-lepton plus jets plus \MET\ and the lepton(s) plus jets plus \MET.

In the 0-lepton search~\cite{:2012rz}, events are selected based on a jet+\MET\ trigger, applying a lepton veto, requiring a minimum number of jets (two to six), $\MET > 160\GeV$, and large azimuthal separation between the \MET\ and reconstructed jets, in order to reject multi-jet background. Events are analysed in five SRs based on jet multiplicity, which are further divided to an overall of eleven channels by using different $m\mathrm{_{eff}(incl.)}$ thresholds. The latter variable is defined as the scalar sum of the transverse momenta of jets with $\pt > 40\GeV$ plus the \MET. The most important sources of background are estimated with data-driven methods, by using measurements in control regions (CRs) and Monte Carlo (MC) predictions for SRs and CRs, applying similar techniques as for the one/two-leptons search described below. The $m\mathrm{_{eff}(incl.)}$ distributions for data, for various background processes before and after fitting to CR observations and for two MSUGRA/CMSSM benchmark model points with $m_0 = 500\GeV$, $m_1/2 = 570\GeV$, $A_0 = 0$, $\tan\beta = 10$ and $\mu > 0$ and with $m_0 = 2500\GeV$, $m_1/2 = 270\GeV$, $A_0 = 0$, $\tan\beta = 10$ and $\mu > 0$ are shown in Fig.~\ref{fig:0lepton} (left).

Limits for squark and gluino production are set in the absence of deviations from SM predictions. Figure~\ref{fig:0lepton} (right) illustrates the 95\% confidence level (CL) limits set under the mSUGRA/CMSSM framework. Exclusion limits are obtained by using the signal region with the best expected sensitivity at each point. In the MSUGRA/CMSSM case, the limit on $m_{1/2}$ reaches 300~\GeV\ at high $m_{0}$ and 640~\GeV\ for low values of $m_0$. Squarks and gluinos with equal masses below 1360~\GeV\ are excluded in this scenario.

\begin{figure}[!htbp]
  \includegraphics[width=0.49\linewidth]{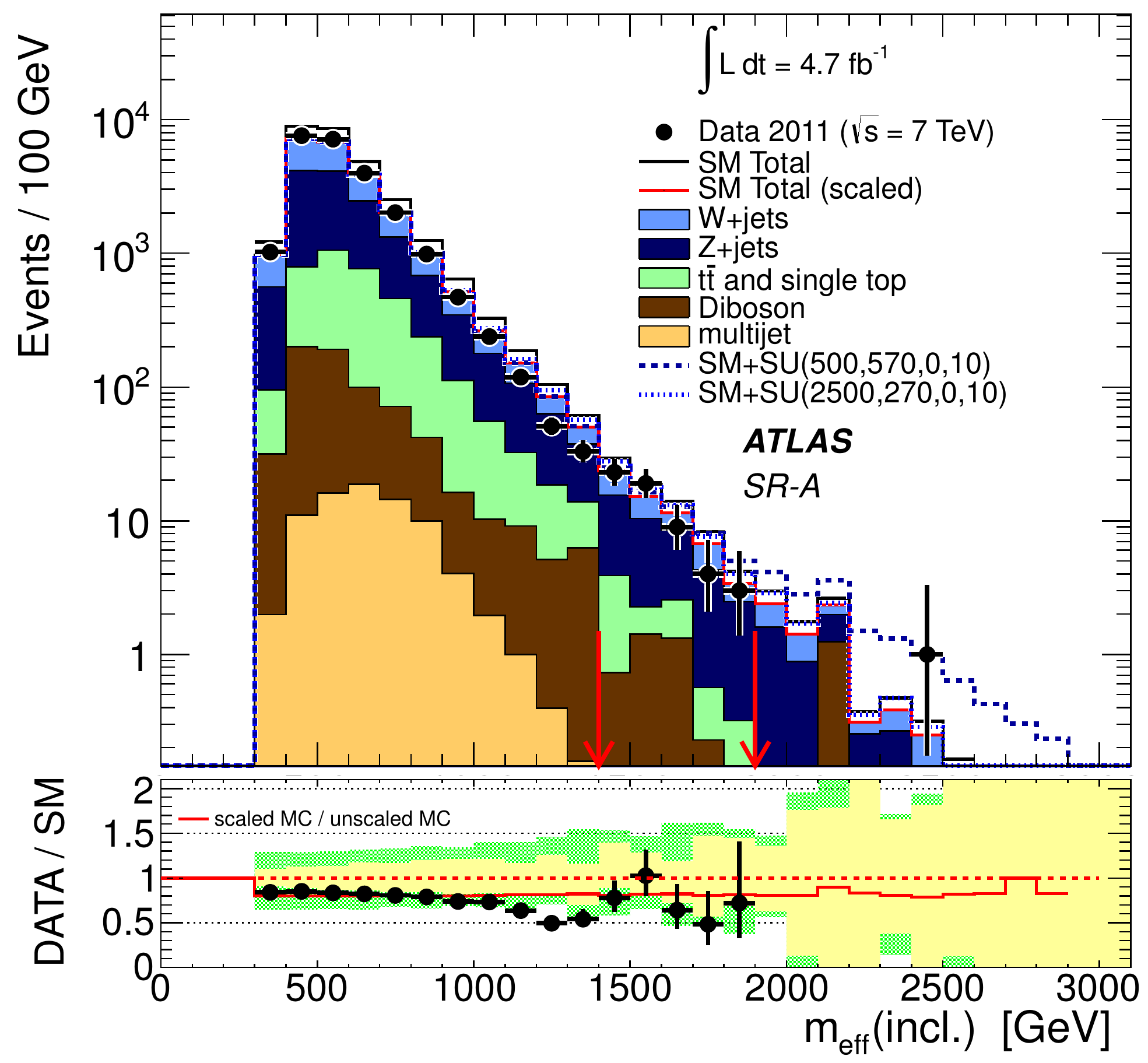} \hfill \includegraphics[width=0.49\linewidth]{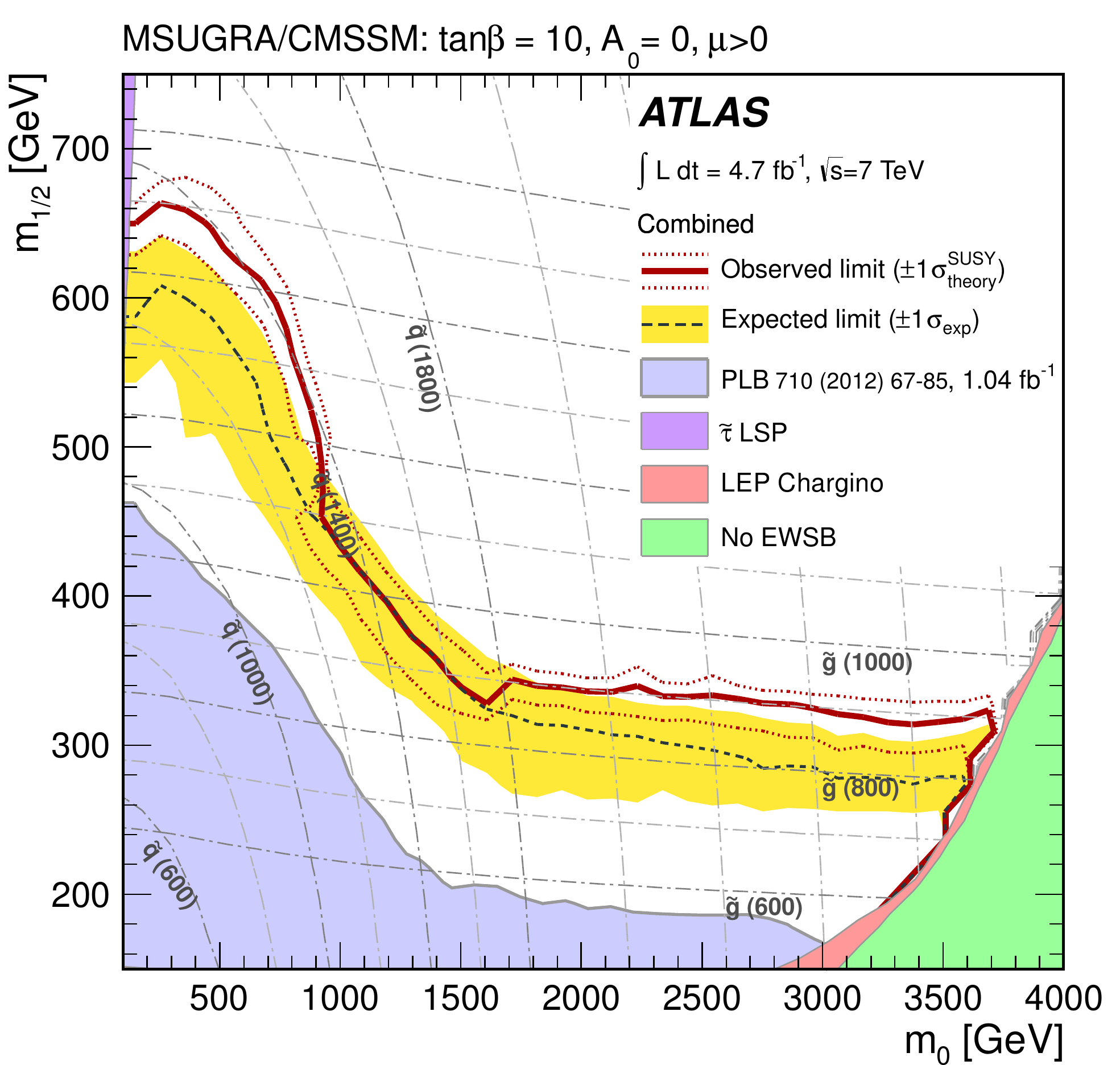} %}
\caption{0-lepton plus jets plus \MET\ analysis. {\it Left:} Observed $m\mathrm{_{eff}(incl.)}$ distribution for the two-jets plus MET analysis (channel~A)~\cite{:2012rz}. In the top panel, the histograms show the SM background expectations, both before (black open histogram) and after (medium (red) open histogram) use of a fit to scale the expectations to CR observations. The bottom panel shows the fractional deviation of the data from the total unscaled background estimate (black points), together with the fractional deviation of the total scaled background estimate from the total unscaled background estimate (medium (red) line). {\it Right:} The 95\% CL exclusion limits on the ($m_0, m_{1/2}$) plane of MSUGRA/CMSSM for $\tan\beta = 10$, $A_0 = 0$ and $\mu > 0$. The black dashed lines show the expected limits, with the light (yellow) bands indicating the $1\sigma$ excursions due to experimental uncertainties. Observed limits are indicated by medium (maroon) curves, where the solid contour represents the nominal limit, and the dotted lines are obtained by varying the cross section by the theoretical scale and PDF uncertainties.}
\label{fig:0lepton}       
\end{figure}

The one- and two-leptons search~\cite{:2012ms} is motivated by models with SUSY decay chains with intermediate $\tilde{\chi}^0$, $\tilde{\chi}^{\pm}$, $\tilde{\ell}$, for which isolated leptons are a clean signature. The SRs are divided into those which require exactly one lepton and three or four jets and those with at least two leptons and two or four jets. Within this search, ATLAS also performs a soft-lepton analysis which enhances the sensitivity of the search in the difficult kinematic region where the neutralino and gluino masses are close to each other forming the so-called ``compressed spectrum''. Further cuts are applied on \MET, $m_{\mathrm{T}}$, $m\mathrm{_{eff}(incl.)}$ and $\MET/m_{\mathrm{eff}}$, where $m_{\mathrm{eff}}$ is defined as the scalar sum of the \MET and the transverse momenta of the selected leptons and jets. 

The major backgrounds ($t\bar{t}$, $W$+jets, $Z$+jets) are estimated by isolating each of them in a dedicated control region, normalising the simulation to data in that control region, and then using the simulation to extrapolate the background expectations into the signal region. The multijet background is determined from the data by a matrix method. All other (smaller) backgrounds are estimated entirely from the simulation, using the most accurate theoretical cross sections available. To account for the cross-contamination of physics processes across control regions, the final estimate of the background is obtained with a simultaneous, combined fit to all control regions.

Apart from other theoretical models (mSUGRA, GMSB), results are also interpreted under simplified-model assumptions. Figure~\ref{fig:12lepton} (left) illustrates the diagrams of two of the topologies used for the interpretation: a one-step $\tilde{q}_L$-pair and a two-step $\tilde{g}$-pair production. In Fig.~\ref{fig:12lepton} (right), the excluded cross sections at 95\% confidence level for the one-step simplified model of gluino pair-production with $\tilde{g} \rightarrow q \bar{q}' \tilde{\chi}_1^{\pm} \rightarrow q \bar{q}' W^{\pm} \tilde{\chi}_1^0$ are shown. The plots are from the combination of the hard and soft single-lepton channels.

\begin{figure}[!htbp]
\begin{minipage}[b]{0.4\linewidth}
  \includegraphics[width=0.88\linewidth]{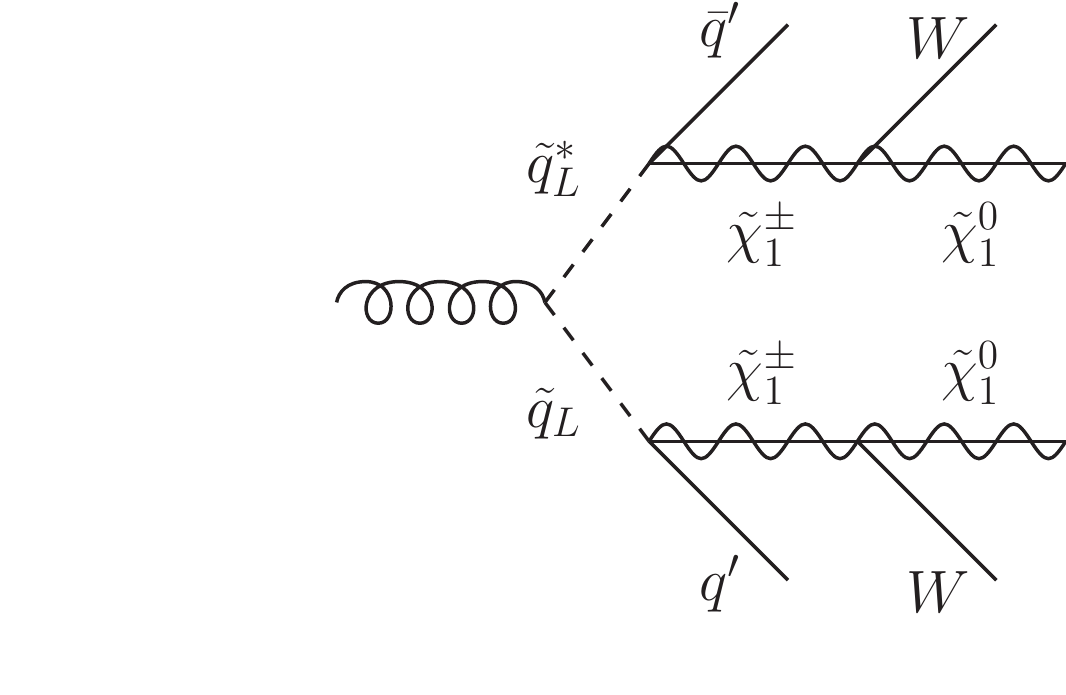}  \includegraphics[width=\linewidth]{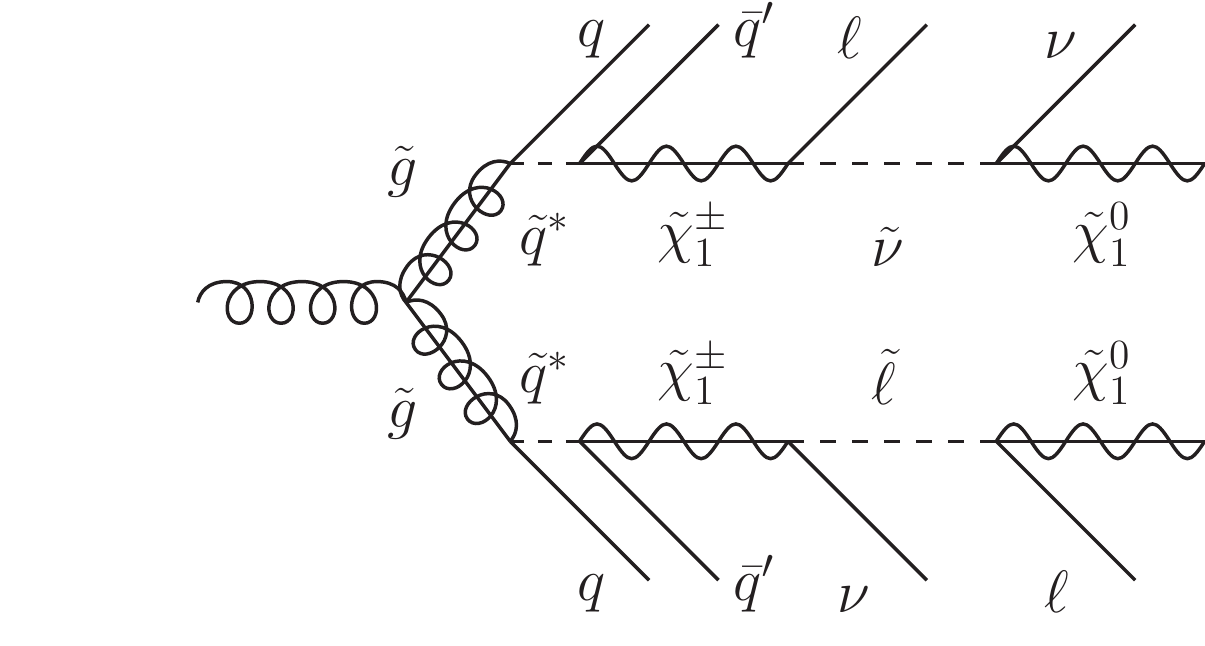}
 \end{minipage}
   \hspace*{1cm} \includegraphics[width=0.49\linewidth]{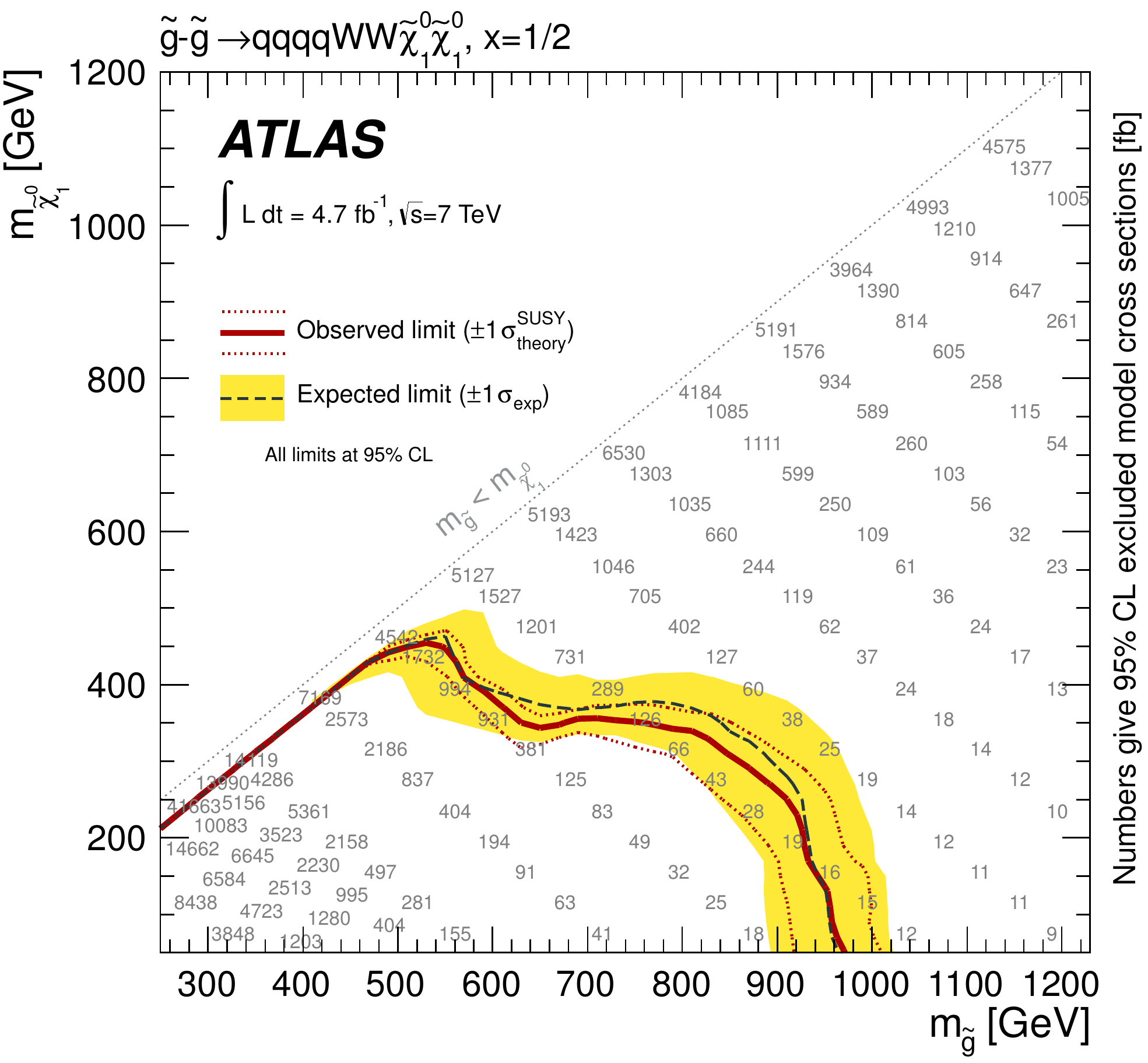} %}
\caption{Leptons plus jets plus \MET\ analysis. {\it Left:} Representative diagrams for different SUSY models: one-step simplified model with $pp \rightarrow \tilde{q}_L \tilde{q}_L^*$ and subsequent decay via charginos (top); two-step simplified model with $pp \rightarrow\tilde{g}\tilde{g}$ and subsequent decays via charginos and sleptons or sneutrinos (bottom).  {\it Right:} Excluded cross sections at 95\% confidence level for the one-step simplified model of gluino pair-production with $\tilde{g} \rightarrow q \bar{q}' \tilde{\chi}_1^{\pm} \rightarrow q \bar{q}' W^{\pm} \tilde{\chi}_1^0$~\cite{:2012ms}. The chargino mass is set to be halfway between gluino and LSP masses. The band around the median expected limit shows the $\pm 1\sigma$ variations on the median expected limit, including all uncertainties except theoretical uncertainties on the signal. The dotted lines around the observed limit indicate the sensitivity to $\pm 1\sigma$  variations on these theoretical uncertainties. The numbers indicate the excluded cross section in femtobarns. }
\label{fig:12lepton}       
\end{figure}

\subsection{Third-generation squarks}
\label{sec:third}

The mixing of left- and right-handed gauge states which provides the mass eigenstates of the scalar quarks and leptons can lead to relatively light 3$^\mathrm{rd}$ generation particles. Stop ($\tilde{t}_1$) and sbottom ($\tilde{b}_1$) with a sub-TeV mass are favoured by the naturalness argument, while the stau ($\tilde{\tau}_1$) is the lightest slepton in many models. Therefore these could be abundantly produced either directly or through  gluino production and decay. Such events are characterised by several energetic jets (some of them $b$-jets), possibly accompanied by light leptons, as well as high \MET.

The first analysis presented here~\cite{:2012pq} comprises the full 2011 dataset of 4.7~\ifb\ and adopts an improved selection that requires large \MET, no electron or muon and at least three jets identified as originating from $b$-quarks ($b$-jets) in the final state. Results are interpreted in simplified models where sbottoms or stops are the only squarks produced in the gluino decays, leading to final states with four $b$-quarks. 

The gluino-sbottom model is an MSSM scenario where the $\tilde{b}_1$ is the lightest squark, all other squarks are heavier than the gluino, and $m_{\tilde{g}} > m_{\tilde{b}_1} > m_{\tilde{\chi}_1^0}$, so the branching ratio for $\tilde{g}\rightarrow\tilde{b}_1 b$ decays is 100\%. Sbottoms are produced via $\tilde{g}\tilde{g}$ or by $\tilde{b}_1\tilde{b}_1$ direct pair production and are assumed to decay exclusively via $\tilde{b}_1\rightarrow b\tilde{\chi}_1^0$, where $m_{\tilde{\chi}_1^0}$ is set to 60~\GeV. Exclusion limits are presented in the $(m_{\tilde{g}},m_{\tilde{b}_1})$ plane in Fig.~\ref{fig:glu-med} (left).

\begin{figure}[!htbp]
  \includegraphics[width=0.49\linewidth]{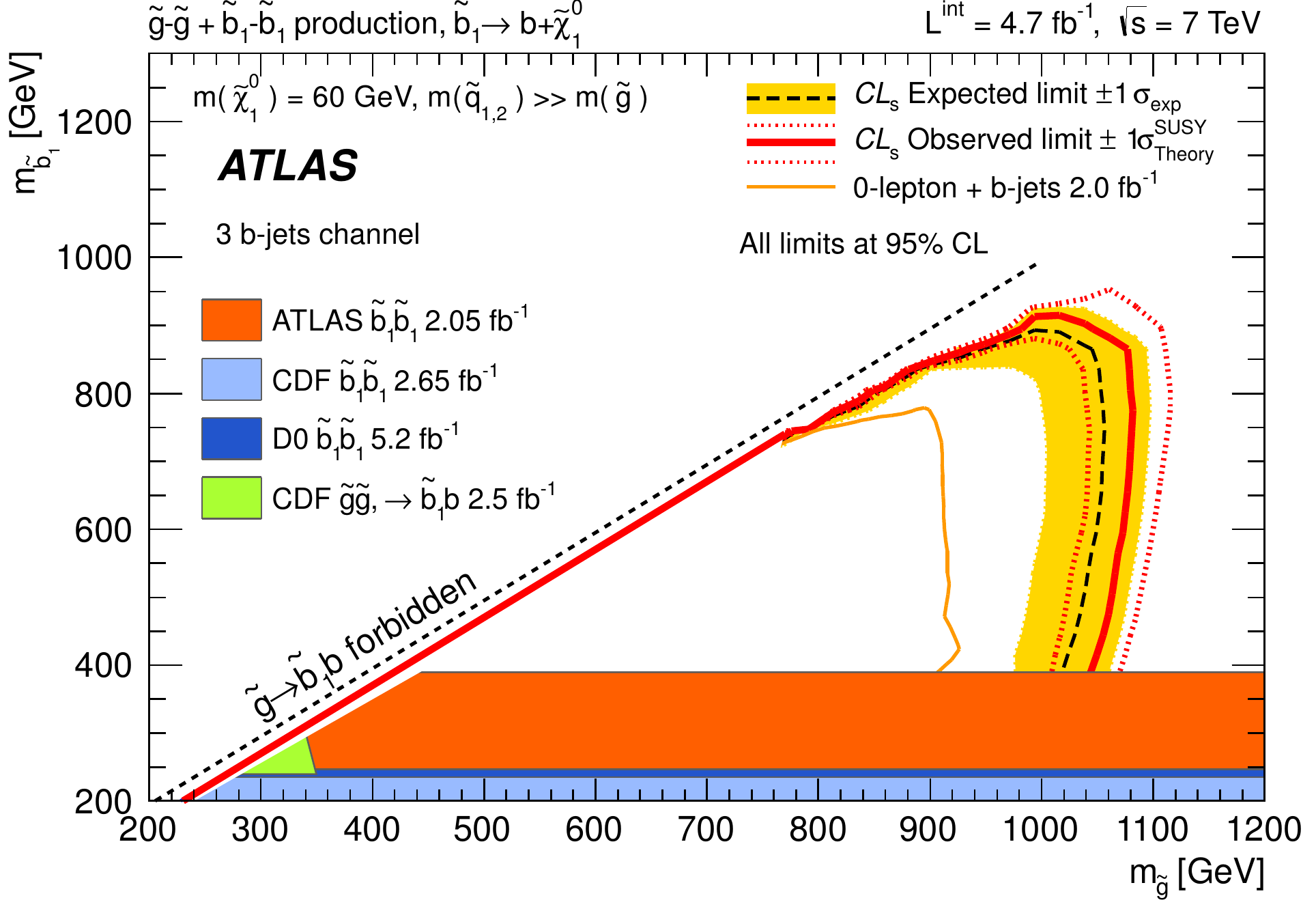} \hfill \includegraphics[width=0.49\linewidth]{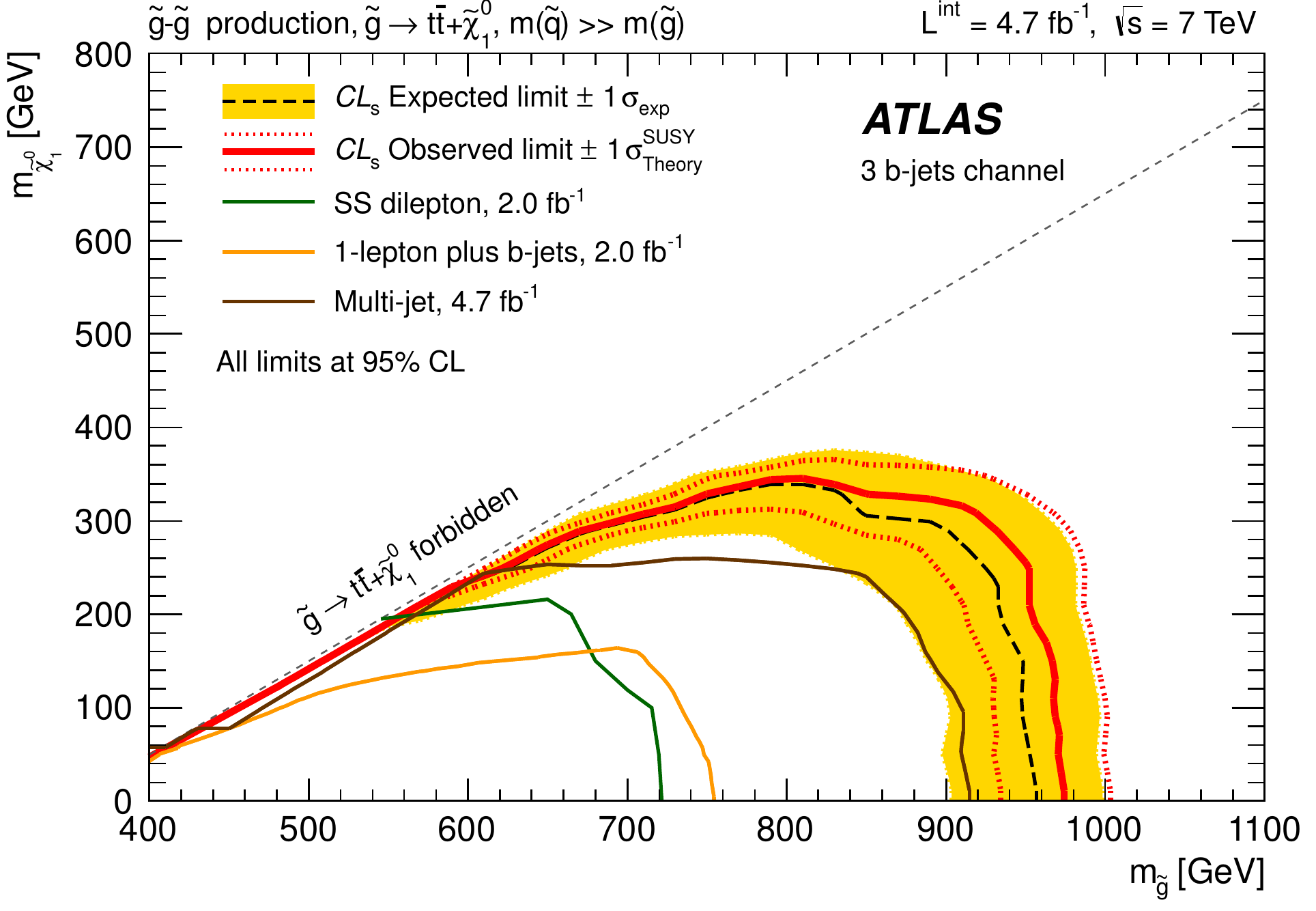} %}
\caption{Results from searches for gluino-mediated third-generation squarks. The dashed black and solid bold red lines show the 95\% CL expected and observed limits respectively, including all uncertainties except the theoretical signal cross-section uncertainty. The shaded (yellow) band around the expected limit shows the impact of the experimental uncertainties while the dotted red lines show the impact on the observed limit of the variation of the nominal signal cross section by $1\sigma$ theoretical uncertainty. {\it Left:} Exclusion limits in the $(m_{\tilde{g}},m_{\tilde{b}_1})$ plane for the gluino-sbottom model~\cite{:2012pq}. Also shown for reference are the previous CDF, D0 and ATLAS analyses.  {\it Right:} Exclusion limits in the $(m_{\tilde{g}},m_{\tilde{\chi}_1^0})$ plane for the Gtt model~\cite{:2012pq}. Also shown for reference are the previous ATLAS analyses.}
\label{fig:glu-med}       
\end{figure}

A simplified scenario (``Gtt model''), where $\tilde{t}_1$ is the lightest squark but $m_{\tilde{g}} < m_{\tilde{t}_1}$, is considered. Pair production of gluinos is the only process taken into account since the mass of all other sparticles apart from the $\tilde{\chi}_1^0$ are above the \TeV\ scale. A three-body decay via off-shell stop is assumed for the gluino, yielding a 100\% BR for the decay $\tilde{g}\rightarrow t\bar{t}\tilde{\chi}_1^0$. The stop mass has no impact on the kinematics of the decay and the exclusion limits are presented in the $(m_{\tilde{g}},m_{\tilde{\chi}_1^0})$ plane in Fig.~\ref{fig:glu-med} (right).

Furthermore, a search for light top squarks has been performed in the dilepton final state: $ee$, $\mu\mu$ and $e\mu$ with 4.7~\ifb~\cite{:2012tx}. The leading lepton \pt\ is required to be less than 30~\GeV, a $Z$-veto is imposed ($|m_{\ell\ell}-m_Z|>10\GeV$) and $\MET>20\GeV$ is required. Good agreement is observed between data and the SM prediction in all three flavour channels. The results are interpreted in the $(m_{\tilde{t}_1}, m_{\tilde{\chi}_1^0})$ plane, shown in Fig.~\ref{fig:light-stop} with the chargino mass set to 106~\GeV, and with the assumption that the decay $\tilde{t}_1\rightarrow b\tilde{\chi}_1^{\pm}$ occurs 100\% of the time, followed by decay via a virtual $W$ ($\tilde{\chi}_1^{\pm}\rightarrow W^{*}\tilde{\chi}_1^0$) with an 11\% branching ratio (per flavour channel) to decay leptonically. A lower limit at 95\% confidence level is set on the stop mass in this plane using the combination of flavour channels. This excludes stop masses up to 130~\GeV\ (for neutralino masses between 1~\GeV\ and 70~\GeV).

\begin{figure}[!htbp]\sidecaption
\resizebox{0.43\hsize}{!}{\includegraphics*{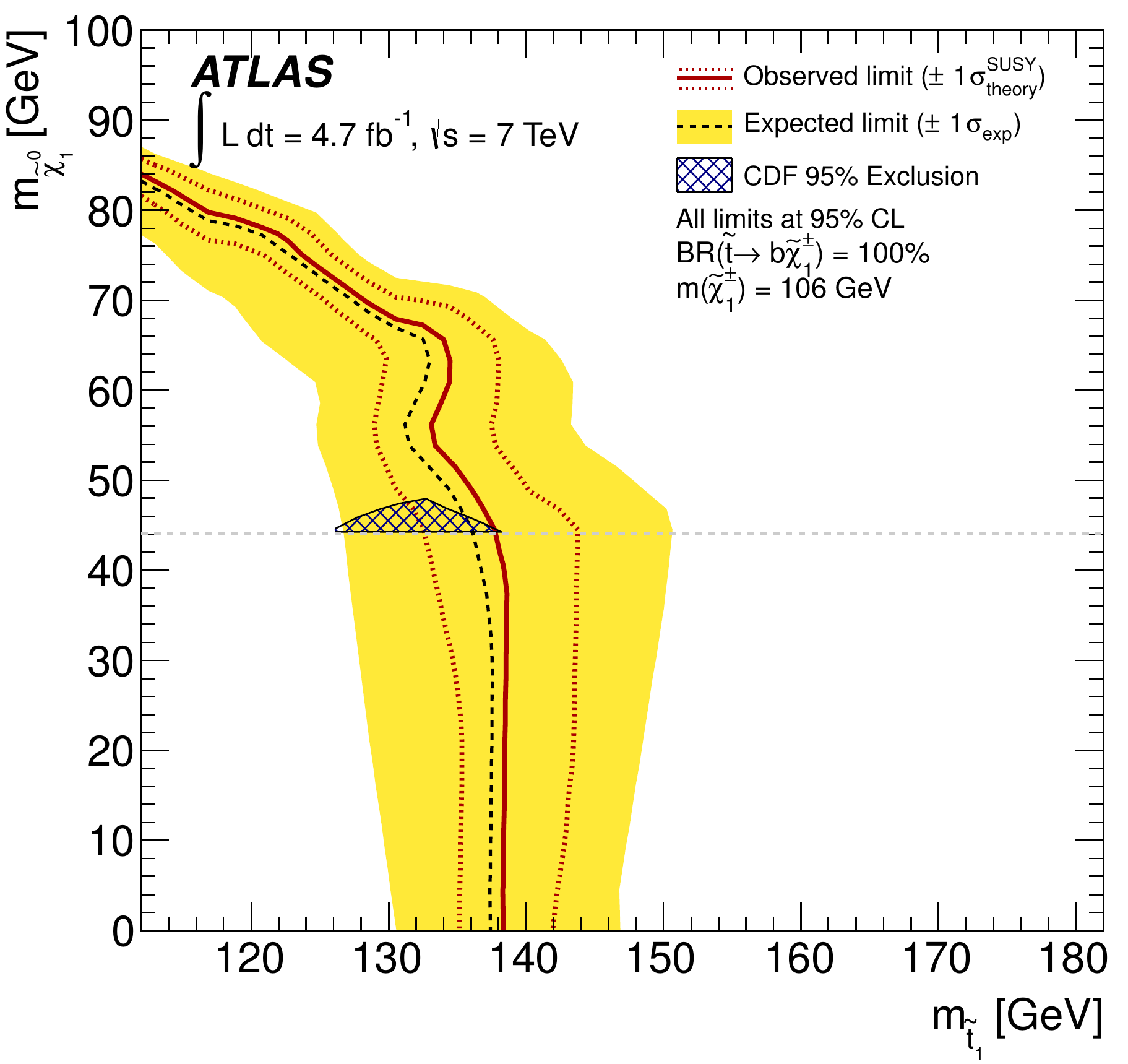}}
\caption{Search for very light stops: 95\% exclusion limit in the $(m_{\tilde{t}_1}, m_{\tilde{\chi}_1^0})$ mass plane, with $m_{\tilde{\chi}_1^{\pm}}=106\GeV$~\cite{:2012tx}. The dashed and solid lines show the 95\% CL expected and observed limits, respectively, including all uncertainties except for the theoretical signal cross-section uncertainty (PDF and scale). The band around the expected limit shows the $\pm 1\sigma$ result. The dotted $\pm 1\sigma$ lines around the observed limit represent the results obtained when moving the nominal signal cross section up or down by the theoretical uncertainty. Illustrated also is the region excluded at the 95\% CL by the CDF experiment, where the lowest neutralino mass considered was 44~\GeV, indicated by the horizontal dotted line. }\label{fig:light-stop}
\end{figure}

\subsection{Direct weak-gaugino production}
\label{sec:gaugino}

Signatures with multiple charged leptons can arise at the LHC through cascade decays of charginos and neutralinos. These weak gauginos can either be produced directly or can result from decays of squarks and gluinos. The analysis presented here~\cite{:2012cwa} consists of a search for direct production of weak gauginos in final states with three leptons and \MET\ at $\rts = 7\TeV$ with 2.06~\ifb.

In one of the theoretical scenarios considered, the phenomenological minimal supersymmetric Standard Model (pMSSM), a series of simplifying assumptions reduces the 105~parameters of the $R$-parity conserving MSSM to~19. These assumptions include no new sources of $CP$ violation and degenerate 1$^\mathrm{st}$ and 2$^\mathrm{nd}$ generation sfermion masses. This analysis made further assumptions, e.g.\ $\tan\beta=6$ to ensure the same leptonic branching fraction for each flavour, to reduce the number of parameters to three: the $U(1)$ gaugino mass $M_1$, the $S\!U(2)$ gaugino mass $M_2$, and the higgsino mass $|\mu|$.

The baseline event selection requires three leptons with $\pt > 10\GeV$, $\MET > 50\GeV$, and at least one same-flavour, opposite-charge (SFOC) lepton pair. Two signal regions have been considered, both vetoing jets identified as originating from $b$-quarks. SR1 is defined by requiring that the invariant mass of the SFOC pair be further than 10~\GeV\ from the $Z$ mass. Conversely, SR2 is defined by requiring the SFOC mass to be within 10~\GeV\ of the $Z$ mass. The SR1 and SR2 selections target SUSY events with intermediate slepton or on-mass-shell $Z$-boson decays, respectively.

\begin{figure}[!htbp]\sidecaption
\resizebox{0.52\hsize}{!}{\includegraphics*{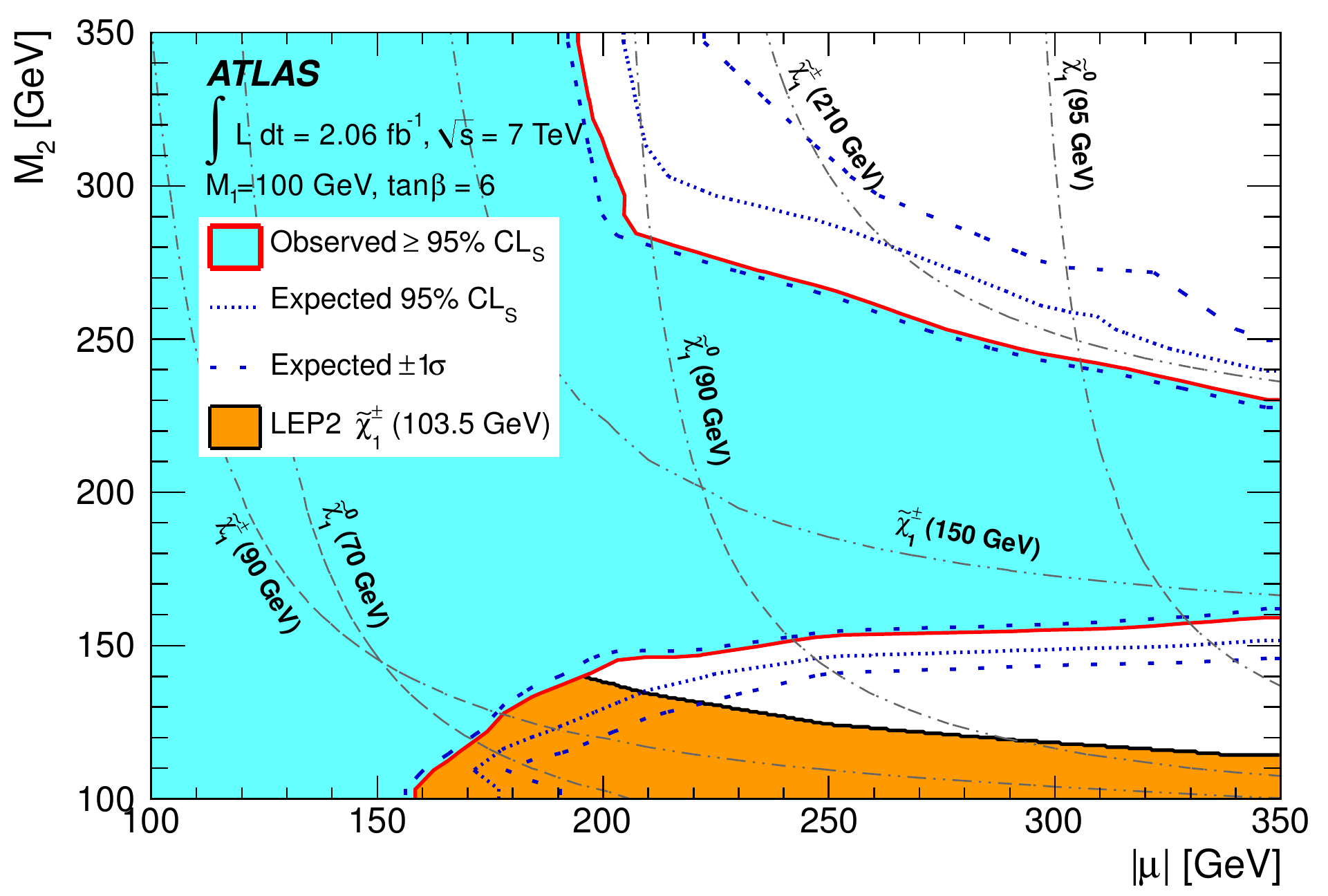}}
\caption{Observed and expected 95\% CL limit contours for chargino and neutralino production in the pMSSM scenario set by the three-leptons and \MET\ channel~\cite{:2012cwa}. }\label{fig:gaugino}
\end{figure}

In SR1 (SR2), 32 (95) events are observed in data. The total SM prediction is $26\pm5$ ($72\pm12$) events. The background-only $p$-value is found to be 19\% (6\%). 95\% CL limits are set on the parameter space of pMSSM, as shown in Fig.~\ref{fig:gaugino}. An upper bound of 9.9~fb (23.8~fb) at 95\% CL has been placed on the visible cross section in SR1 (SR2).

\section{$R$-parity violating SUSY}
\label{sec:RPV}

Searches are also performed in ATLAS for several signatures associated with the violation of $R$-parity (RPV). In one of them, a term $W_\mathrm{RPV} = \lambda'_{ijk}\tilde{u}_j \bar{d}_k l_i$ is introduced into the SUSY Lagrangian, which in turn permits a process $d\bar{d} \rightarrow e^- \mu^+$ via $t$-channel top squark exchange. In 2.1~\ifb\ of data, ATLAS has performed the first search for continuum production of a muon and electron of opposite sign~\cite{Aad:2012yw}, finding no excess above the SM expectation. As demonstrated in Fig.~\ref{fig:RPV} (left), for a coupling parameter product of $|\lambda'_{131}\lambda'_{231}|=|\lambda'_{132}\lambda'_{232}|=0.05$, such processes are ruled out at 95\% CL for $m_{\tilde{t}} < 200\GeV$.

\begin{figure}[!htbp]
  \includegraphics[width=0.46\linewidth]{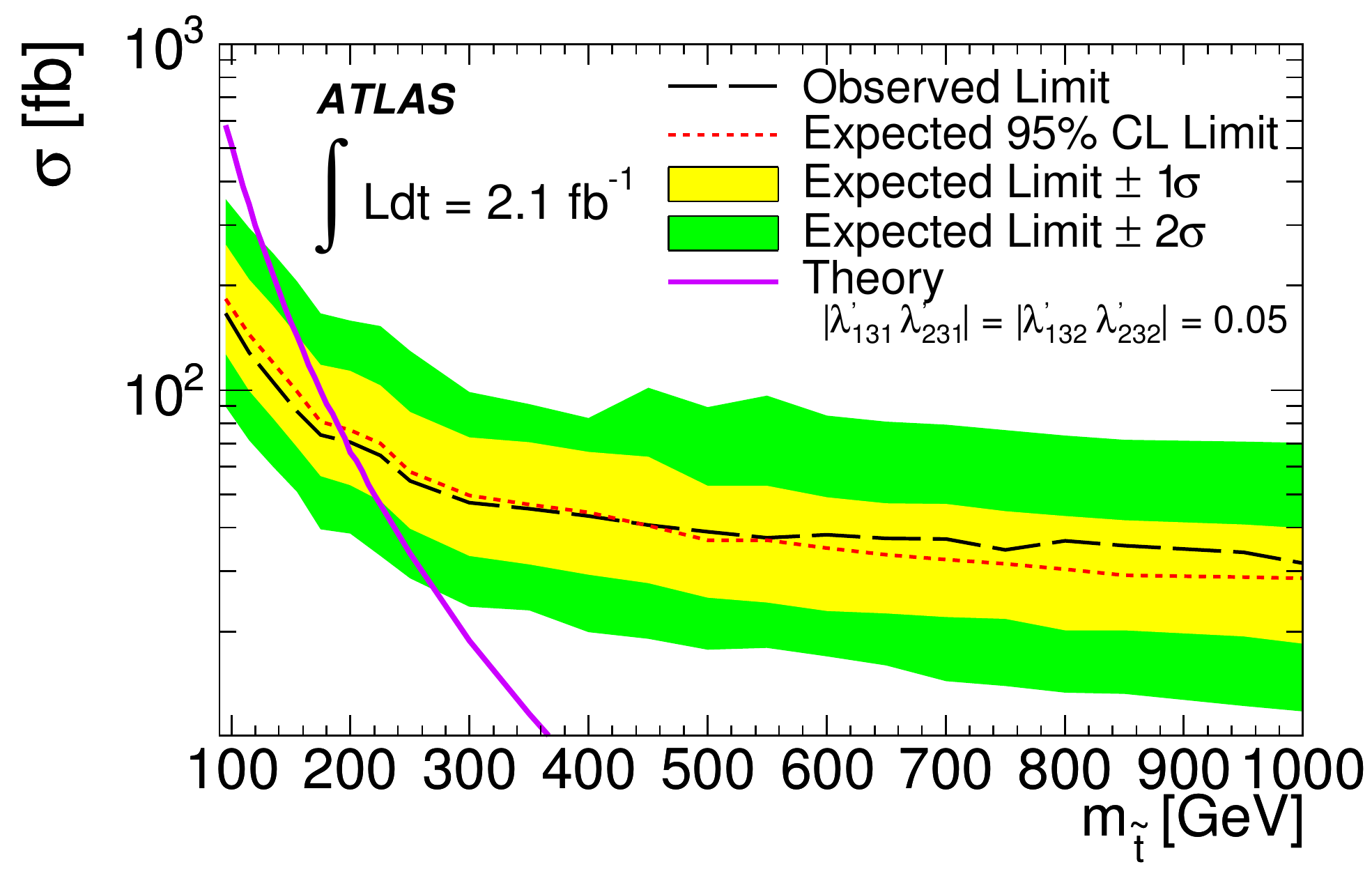} \hfill \includegraphics[width=0.52\linewidth]{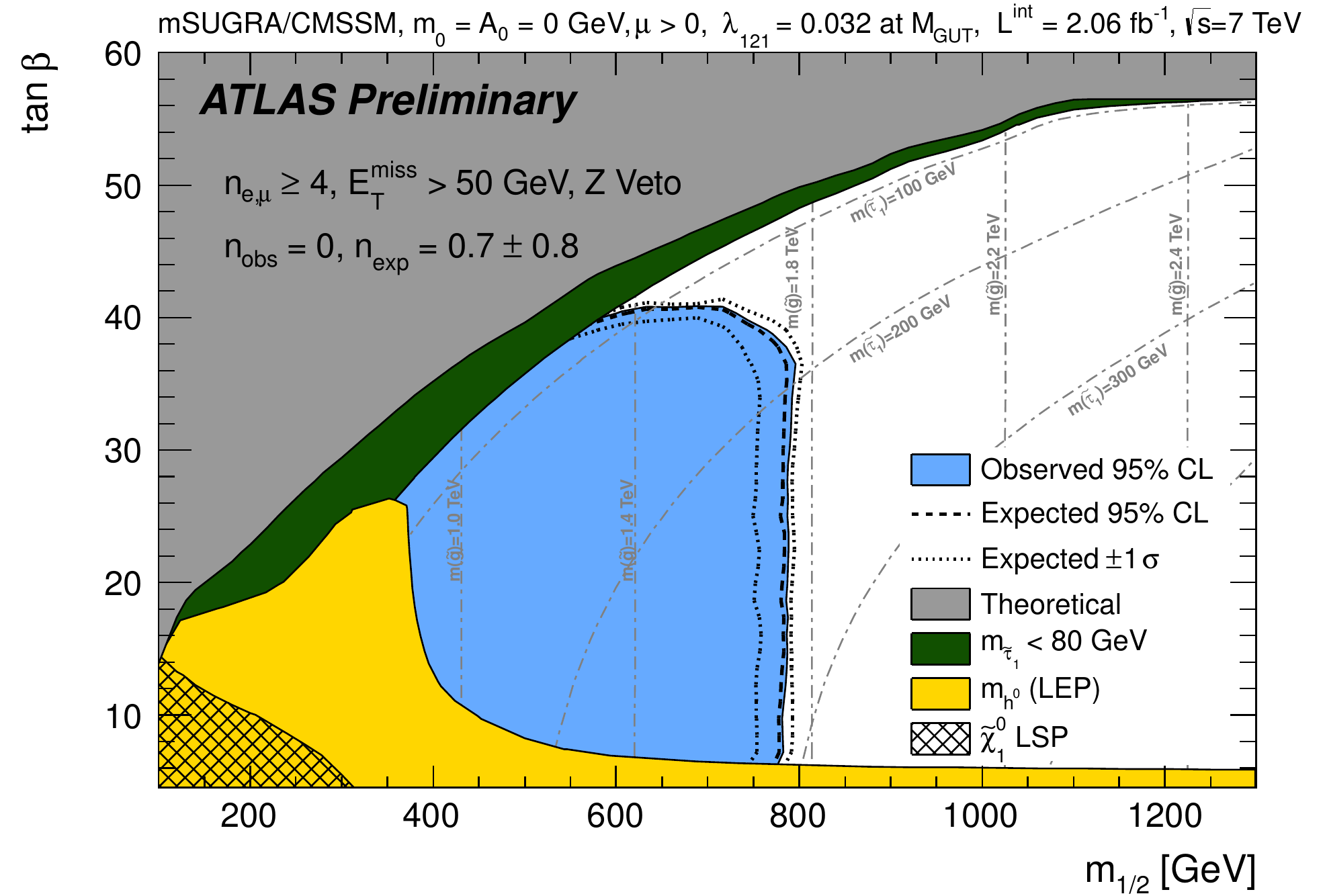} %}
\caption{{\it Left:} The observed 95\% CL upper limits on $\sigma(pp \rightarrow e\mu)$ through the RPV $t$-channel exchange of a scalar top quark as a function of $m_{\tilde{t}}$~\cite{Aad:2012yw}. The expected limits are also shown together with the $\pm 1$ and $\pm 2$ standard-deviation uncertainty bands. The theoretical cross section for $|\lambda_{131}^{\prime}\lambda_{231}^{\prime}|=|\lambda_{132}^{\prime}\lambda_{232}^{\prime}|=0.05$ is also shown. {\it Right:} Excluded region at 95\% CL as a function of $m_{1/2}$ and $\tan\beta$ obtained with the four-leptons analysis~\cite{stauLSP}. The expected exclusion and its $\pm 1\sigma$ variations are indicated by dashed lines. The other solid shaded areas are excluded from this analysis by LEP results on the Higgs mass or because $m_{\tilde{\tau}_1} < 80\GeV$.   }
\label{fig:RPV}       
\end{figure}

A search requiring four or more leptons (electrons or muons) in the final state~\cite{4leptons} is sensitive to various supersymmetric models including pair-production of strongly interacting SUSY particles with $R$-parity breaking decays of a $\tilde{\tau}_1$ LSP~\cite{stauLSP}. Moderate missing transverse momentum is expected in the final state due to the presence of neutrinos originating in the decay of the LSP. 

Isolated electrons (muons) with $\pt > 10\GeV$ and pseudorapidity $|\eta| < 2.47$ ($|\eta| < 2.4$) are considered. A signal region selecting events with at least four leptons, $\MET> 50\GeV$ and a veto on events containing a $Z$-boson candidate is defined. At least one of the selected leptons has to be in the efficiency plateau ($\pt^{e} >25\GeV$ and $\pt^{\mu} >20\GeV$) and match a lepton firing the trigger.

With 2.06~\ifb\ of $pp$ collision data, zero events are observed, while $0.7\pm0.8$ events are expected from SM processes. Observed (expected) upper limits of 1.5 (1.5)~fb set on the visible cross-sections for new phenomena are subsequently used to constrain the mSUGRA/CMSSM scenario with $m_0=A_0=0$, $\mu > 0$, and one $R$-parity lepton flavour violating parameter $\lambda_{121}=0.032$ at $m_\mathrm{GUT}$. In this scenario, the RPV coupling is small enough so that the SUSY particle pair production dominates, and large enough that the $\tilde{\tau}_1$ LSP decays promptly. Values of $m_{1/2} <800\GeV$ are excluded at 95\% CL if $\tan\beta <40$ and $m_{\tilde{\tau}_1} >80\GeV$, as demonstrated in Fig.~\ref{fig:RPV} (right). These are the first limits from the LHC experiments on a model with a $\tilde{\tau}_1$ as the lightest supersymmetric particle.

\section{Meta-stable particles}
\label{sec:LL}

We discuss here the results of a search for the decay of a heavy long-lived particle producing a multi-track displaced vertex (DV) that contains a high-\pT\ muon at a distance between millimeters and tens of centimeters from the $pp$ interaction point~\cite{Aad:2011zb}. The results are interpreted in the context of an RPV SUSY scenario, where such a final state occurs in the decay $\tilde{\chi}_1^0\rightarrow\mu q\bar{q}' $, allowed by the non-zero RPV coupling $\lambda'_{2ij}$.

Events that pass a single-muon trigger of $\pt^{\mu} > 40\GeV$ are selected. The  reconstruction of a DV begins with the selection of high-impact-parameter tracks with $\pt > 1\GeV$. At least four tracks in the DV are required, to suppress background from random combinations of tracks and from material interactions. Background due to particle interactions with material is further suppressed by requiring $m_\mathrm{DV} > 10\GeV$, where $m_\mathrm{DV}$ is the invariant mass of the tracks originating from the DV. High-$m_\mathrm{DV}$ background arise from random spatial coincidence of such a low-$m_\mathrm{DV}$ vertex with a high-\pt\ track is suppressed by vetoing vertices that are reconstructed within regions of high-density material.

Figure~\ref{fig:DV} (left) shows the distribution of $m_\mathrm{DV}$ versus $N\mathrm{^{trk}_{DV}}$ for vertices in the selected data events, including vertices that fail the requirements on $m_\mathrm{DV}$ and $N\mathrm{^{trk}_{DV}}$, overlaid with the signal distribution for a signal sample with $m_{\tilde{q}} = 700\GeV$ and $m_{\tilde{\chi}_1^0} = 494\GeV$. Fewer than 0.03 background events are expected in the data sample of 33~\ipb, and no events are observed. Based on this null observation, upper limits are set on the supersymmetry production cross-section $\sigma\times B$ of the simulated signal decay chain for different combinations of squark and neutralino masses and for different values of $c\tau$, where $\tau$ is the neutralino lifetime (cf.\ Fig.~\ref{fig:DV}, right).

\begin{figure}[!htbp]
  \includegraphics[width=0.48\linewidth]{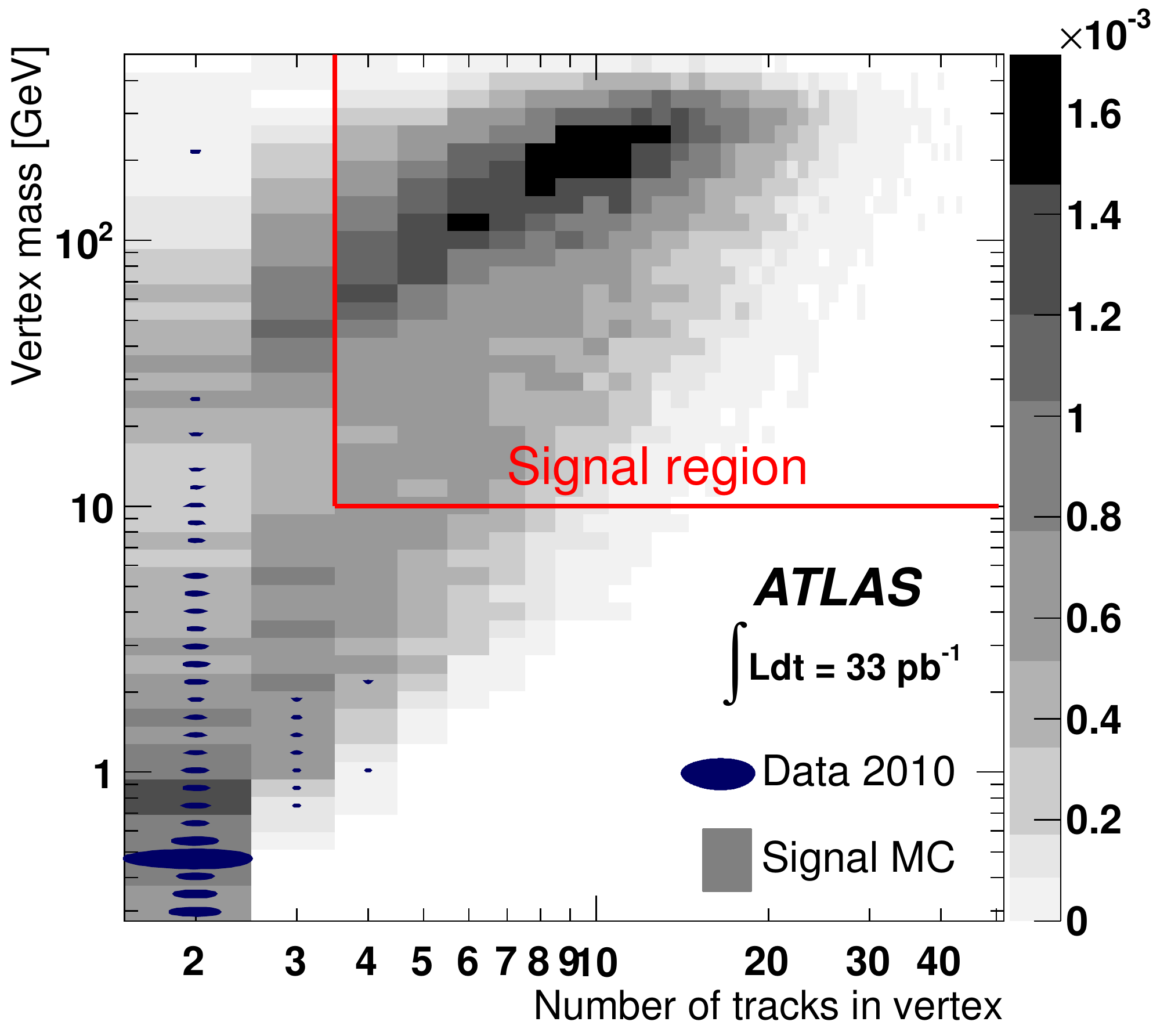} \hfill \includegraphics[width=0.48\linewidth]{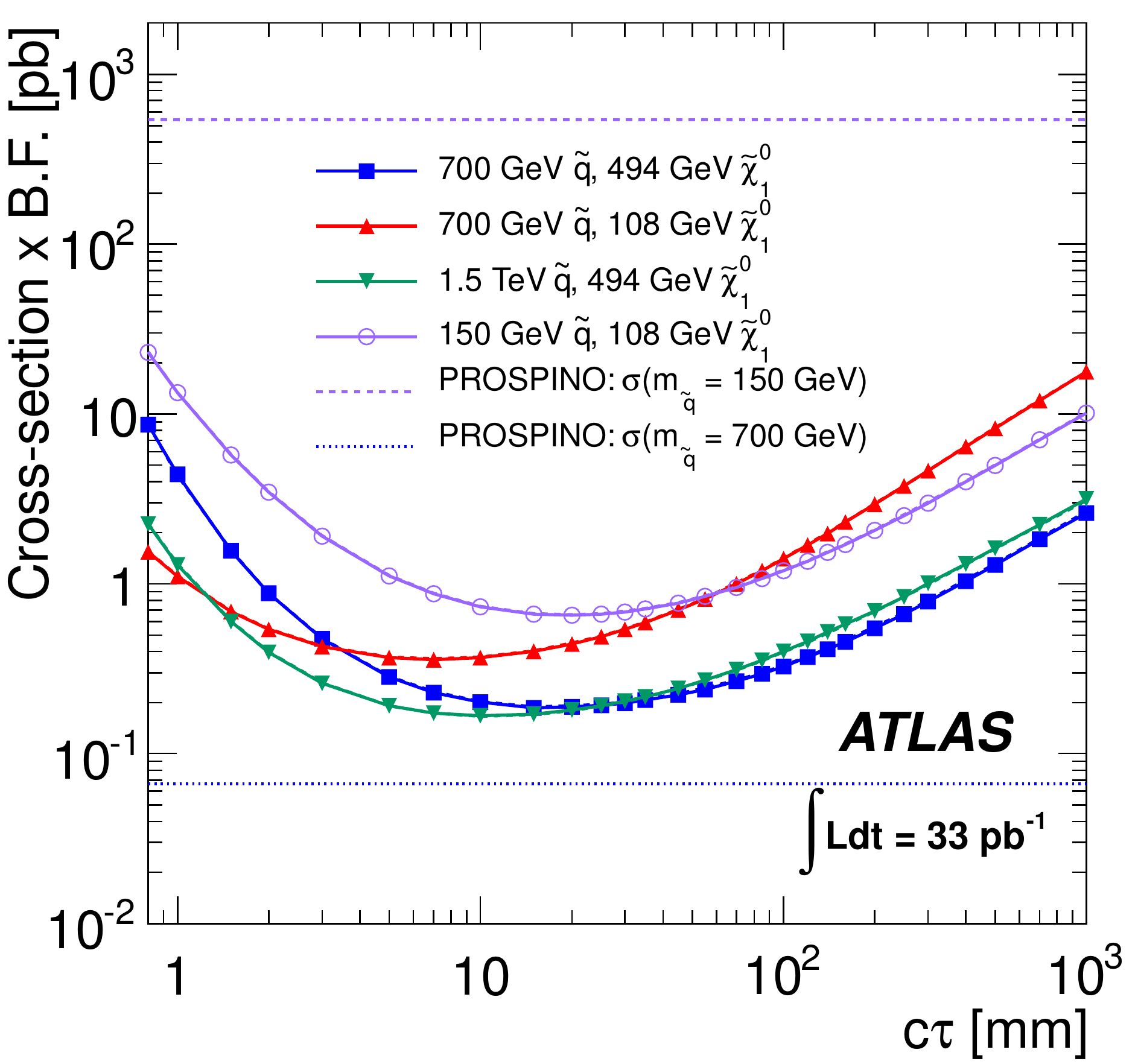} %}
\caption{{\it Left:} Vertex mass ($m_\mathrm{DV}$) versus vertex track multiplicity ($N^\mathrm{trk}_\mathrm{DV}$) for displaced vertices that pass the event selection requirements except the $m_\mathrm{DV}$ and $N^\mathrm{trk}_\mathrm{DV}$ requirements, which are not applied~\cite{Aad:2011zb}. Shaded bins show the distribution for signal MC, and data are shown as filled ellipses, with the area of the ellipse proportional to the number of events in the corresponding bin. {\it Right:} Upper limits at 95\% CL on the production cross-section times branching fraction versus the neutralino proper decay length for different combinations of squark and neutralino masses, based on the observation of zero events satisfying all criteria in a 33~\ipb\ data sample~\cite{Aad:2011zb}. The horizontal lines show the cross-sections calculated from PROSPINO for squark masses of 700~\GeV\ and 150~\GeV. }
\label{fig:DV}       
\end{figure}

In another search for long-lived particles, SUSY is looked for through disappearing tracks, motivated by anomaly-mediated symmetry breaking (AMSB) models where the chargino can live long enough to be detected within the inner detector volume. Since the chargino and the neutralino are almost degenerate in mass in these models, the charged particle ($\pi^{\pm}$) from the decay of this chargino is too soft to be reconstructed, therefore a disappearing track is expected. Events are selected based on large \met, high jet multiplicity and a lepton veto. Chargino candidates are selected among good-quality tracks before the TRT (outer part of the inner detector with a radius between 56 to 108~cm) and less than five hits in the TRT outer module. A comparison of the number of hits in the TRT outer volume between the signal, the SM background and the data is presented in Fig.~\ref{fig:kinked} (left). Constraints on the AMSB chargino mass and lifetime were set with 1.02~\ifb; a chargino having $m_{\tilde{\chi}_1^{\pm}} < 92\GeV$ and $0.5 < \tau_{\tilde{\chi}_1^{\pm}} < 2~\mathrm{ns}$ was excluded at 95\% CL, as illustrated in Fig.~\ref{fig:kinked} (right).

\begin{figure}[!htbp]
  \includegraphics[width=0.48\linewidth]{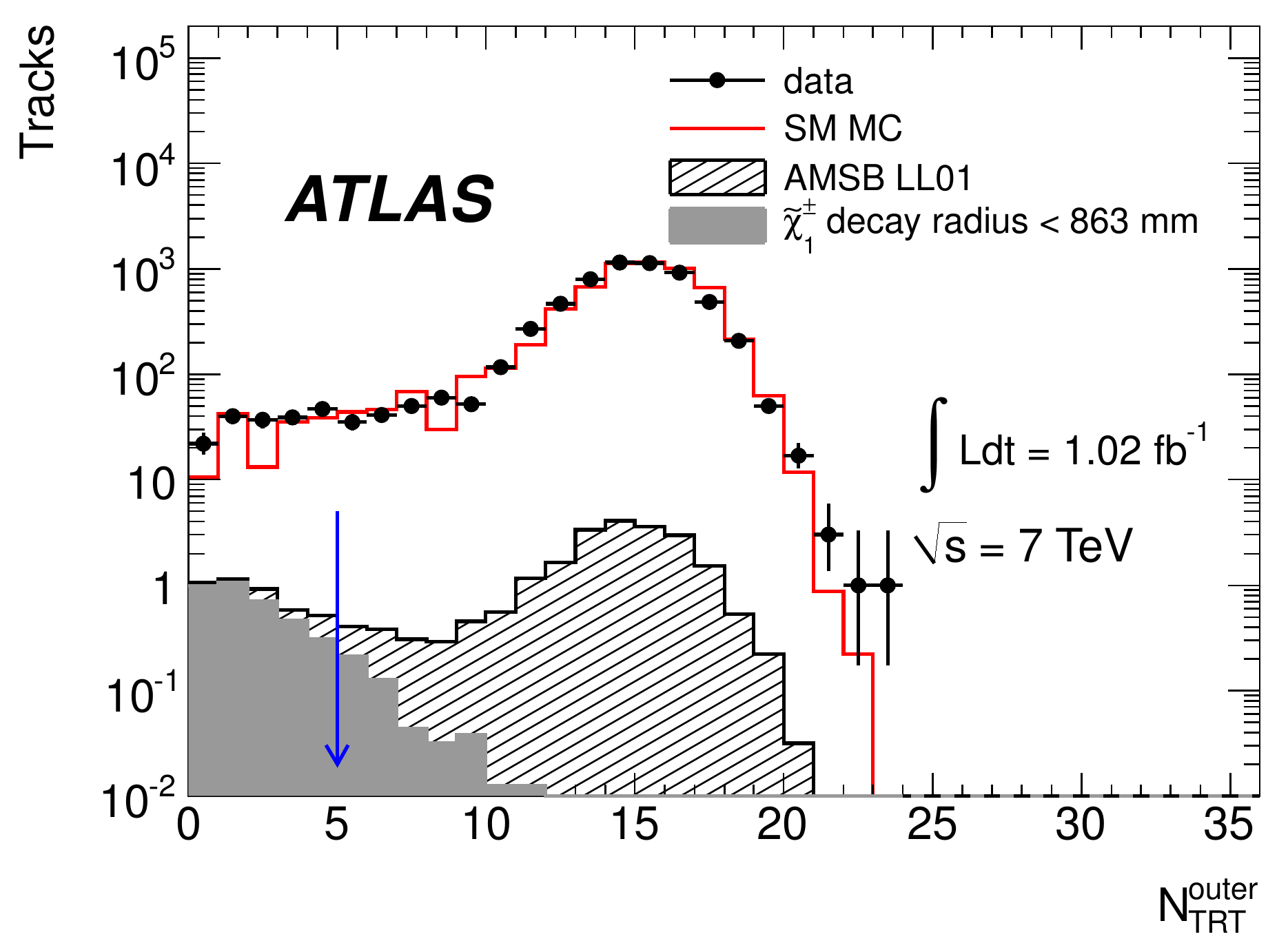} \hfill \includegraphics[width=0.48\linewidth]{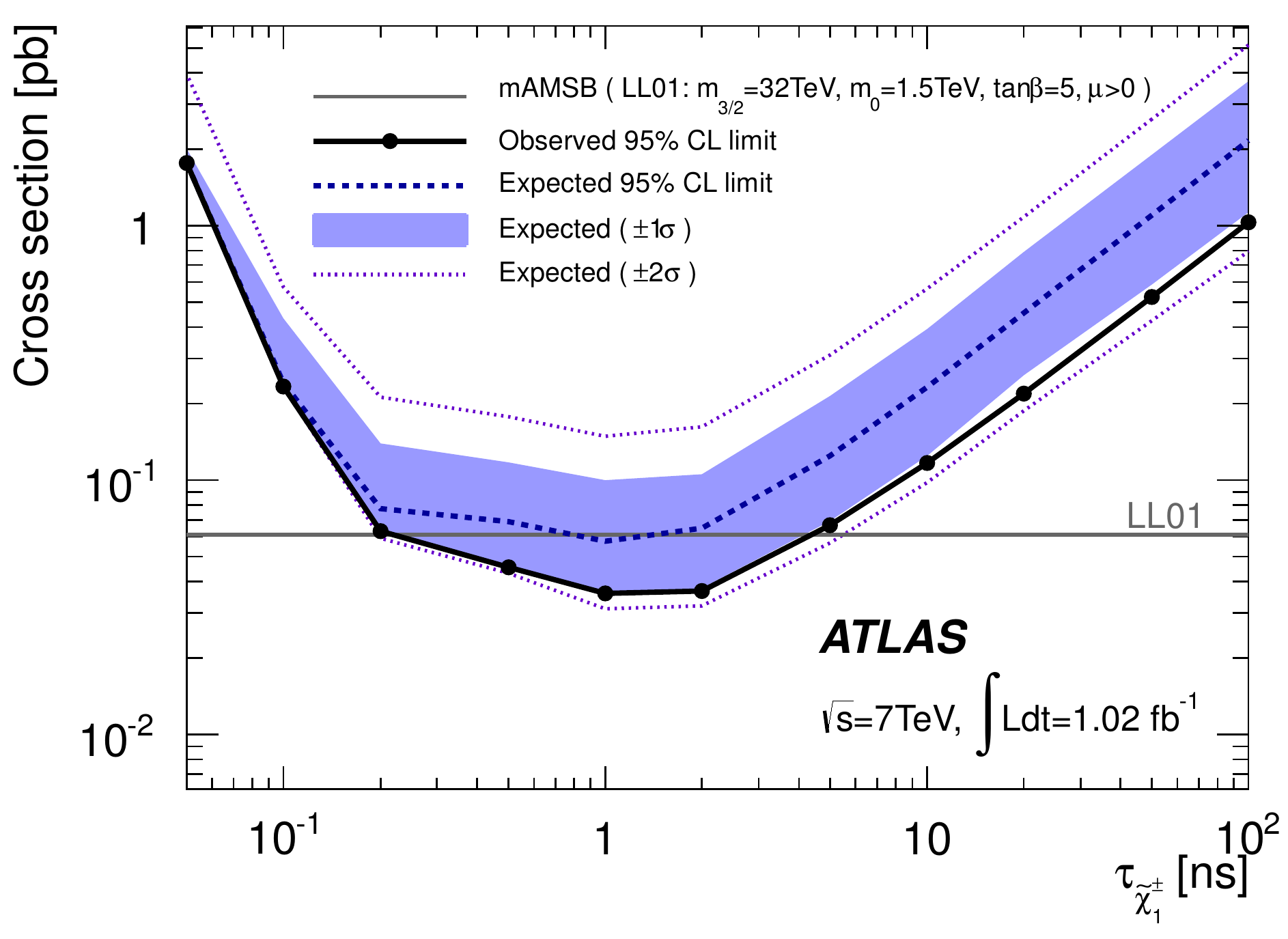} %}
\caption{{\it Left:} The number of hits in the TRT outer module ($N_\mathrm{TRT}^\mathrm{outer}$) for data and LL01 signal events ($\tau_{\tilde{\chi}_1^{\pm}} \simeq 1$~ns) with the high-\pt\ isolated track selection~\cite{ATLAS:2012ab}. The selection boundary is indicated by the arrow. The expectation from SM MC events, normalised to the number of observed events, is also shown. When charginos decay before reaching the TRT outer module, $N_\mathrm{TRT}^\mathrm{outer}$ is expected to have a value near zero; conversely, SM charged particles traversing the TRT typically have $N_\mathrm{TRT}^\mathrm{outer} \simeq 15$. {\it Right:} The observed and expected 95\% CL upper limits on the signal cross section as a function of chargino lifetime for $m_\mathrm{chargino} = 90.2\GeV$~\cite{ATLAS:2012ab}. The bands indicate the $\pm1\sigma$ and $\pm2\sigma$ variations on the median expected limit (dotted line) due to uncertainties. }
\label{fig:kinked}       
\end{figure}

\section{Summary}
\label{sec:summ}

Supersymmetry signals have been sought after by the ATLAS experiment, motivated by various models and topologies: strong production, $3^\mathrm{rd}$ generation fermions, mass degeneracies, $R$-parity violation, among others. They lead to a wide spectrum of signatures: \MET + jets + leptons / photons / $b$-jets / $\tau$-leptons, displaced vertices, not possible to cover all of them here; analyses based on photons and $\tau$-leptons are detailed in Refs.~\cite{ATLAS:2012ag,Aad:2012rt} and~\cite{Aad:2011zj}, respectively.  No deviation from known SM processes has been observed so far with $\sim 5~\ifb$ at $\rts = 7\TeV$. As both techniques and strategy keep evolving, ATLAS will keep looking for supersymmetry with the new data that become available at the LHC.
 
\begin{acknowledgement}
The author acknowledges support by the Spanish MINECO under the project FPA2009-13234-C04-01 and by the Spanish Agency of International Cooperation for Development under the PCI project A1/035250/11.
\end{acknowledgement}

\end{document}